\documentstyle[12pt]{article}
\begin{document}
\renewcommand{\thepage}{ }
\begin{titlepage}
\title{
\hfill
\vspace{1.5cm}
{\center Monopoles in the gauge theory of the t-J model}
}
\author{
R. M\'elin, B. Dou\c{c}ot\\
{}\\
{CRTBT-CNRS, 38042 Grenoble BP 166X c\'edex France}}
\date{}
\maketitle
\begin{abstract}
\normalsize
We consider a RG approach for the plasma
of magnetic monopoles of the
Ioffe-Larkin approach to the
t-J model. We first derive the interaction
parameters of the 2+1 plasma
of magnetic monopoles. The total
charge along the time axis is
constrained to be zero for each
lattice plaquette.
Under the one-plaquette approximation,
the problem is
equivalent to a one dimensional
neutral plasma interacting via a potential
$V(t) \sim t^{\alpha}$,
with $\alpha=1/3$. The plasma
is in a dipolar phase if
$\alpha \ge 1$ and a possibility of
transition towards a Debye screening phase
arises if $\alpha < 1$, so that
there exists a critical Fermi
wave vector $k_f^{*}$ such as
the plasma is Debye screening
if $k_f<k_f^{*}$ and confined if
$k_f>k_f^{*}$.
The 2+1 dimensional problem is
treated numerically. We show
that $k_f^{*}$ decreases and goes
to zero as the number of colors increases.
This suggests that the assumption
 of spin-charge decoupling
within the slave-boson scheme is
self-consistent at large
enough values of $N$ and small
enough doping.
Elsewhere, a confining force
between spinons and antiholons
appears, suggesting a transition
to a Fermi liquid state.
\end{abstract}
\end{titlepage}

\newpage
\renewcommand{\thepage}{\arabic{page}}
\setcounter{page}{1}
\baselineskip=17pt plus 0.2pt minus 0.1pt
\tableofcontents
\section{Introduction}
The physics of strongly interacting
electron systems has received
considerable attention over the
recent years, and it still
bears many challenges, especially on
the theoretical side.
Among the various methods and ideas
which have been explored
in this context, gauge theories seemed
to offer a rather attractive
approach \cite{Ref1}-\cite{Ref7}.
Their essence is to focuss on the
presence of a non double occupancy
constraint, which leads to a local
$U(1)$ symmetry if a slave
boson representation is used
\cite{Ref1}-\cite{Ref3}.
Although it is possible to derive
these gauge theories from
an expansion around a large $N$ mean
field theory of the t-J model
\cite{Ref4} \cite{Ref5}, they could
also be regarded as promising
candidates for an effective low-energy
theory in order to describe
for instance the anomalous normal state
properties of high-$T_c$
superconductors. They seem to predict a
phase diagram for the
single band t-J model which qualitatively
ressembles the experimental
ones for copper-oxide superconductors \cite{Ref6}.
Further more, they reconcile the existence
of a large Fermi surface
corresponding to Luttinger's theorem,
as shown by photoemission
experiments, and the anomalous transport
properties, which are
mostly governed by holes \cite{Ref6}
\cite{Ref7}. Thermodynamic
properties have also been investigated,
and a good agreement
with high temperature expansions for
the t-J model has been reached
\cite{Ref8}. However, this work has
also pointed out that fluctuations
of the gauge field are large, in the
sense that the
variance of the local
statistical flux around a given plaquette is not small
in units of
$2 \pi$, even down to low temperatures. This feature
suggests that the
presence of the lattice may not be inessential, since
it induces a periodic
action as a function of the time and space dependent
flux per plaquette.
As demonstrated by Poliakov, this periodic nature of
the gauge field
has dramatic consequences on 2+1 dimensional electrodynamics
since it allows
for non trivial space time configurations of the gauge field
(monopoles),
which induce charge confinement \cite{Ref9}.
It should be emphasized that in the
context of the t-J model,
the gauge field Lagrangian density is
not the usual $-\frac{1}{4}
F_{\mu \nu} F^{\mu \nu}$ term, but it is
generated upon integrating
out fermionic and bosonic fluctuations
\cite{Ref3}.
A perturbative estimate of a single
monopole action has also been
derived in \cite{Ref3}, and was found
to diverge. However, it is
clear that some globally neutral
configurations (i.e. with the same
number of instantons and anti-instantons)
have a finite action, and
the main question is whether the
corresponding two-component plasma
exhibits Debye-screening or not.
This viewpoint has been developed by
Nagaosa \cite{Ref10} where he
assumed a dissipative-type action for
the gauge field, which
may be relevant for the t-J model at
small temperatures, since it
requires a finite dc conductivity for
the fermions.
His main conclusion is that no major
instanton effect is present
in the t-J model range of parameter
since the dissipative
nature of the gauge field dynamics
strongly inhibits
quantum tunneling.
In the present paper, we address this question from a
slightly different
perspective, with emphasis on the possible zero
temperature transitions.
By contrast to the results of reference \cite{Ref10},
we find that
assuming a Ioffe-Larkin form for the monopole plasma
action leads to a
phase transition between a Debye screening phase
(which corresponds
to a confining force between spinons and antiholons),
and a dipole phase
(leading to unconfined spinons and antiholons).
The control parameters are
the band filling of the underlying t-J model,
and the number $N$
of fermion colors (the physical case being $N=2$).
In rather good
agreement with physical intuition, the dipole phase
of the plasma is found
at large $N$ and small doping. In this regime,
spin-charge separation
may then be a self-consistent hypothesis.
We note however that
this leads either to a renormalized Fermi
liquid or an anomalous
liquid depending on whether Bose condensation of
holons occurs or not.
In the other phase, the gauge field
cannot be treated perturbatively,
and the corresponding mesons (bound
states of spinons and antiholons) are
physical electrons. By contrast, a
transition to the dipole phase
is obtained in reference \cite{Ref10} in the
 presence of an infinitesimal
dissipation. The main difficulty
of the problem
is the determination
of the phase diagram of the monopole
plasma which exhibits long
range interactions (in space and
imaginary time).
Furthermore, unlike the case of the
standard compact
2+1 electrodynamics,
a strong anisotropy exists between
space and time directions,
reflecting the lack of Lorentz
invariance in the model.
This reduced symmetry increases the
difficulty of real space RG
analysis since the functional form
of the monopole interaction
potential is not stable under a RG
transformation. Our approach
has attempted to take advantage of
the fact that the interaction
is much stronger
along the time direction, with a
$\tau^{1/3}$ dependence. The
corresponding one dimensional problem exhibits
a phase transition between
a Debye-screening phase and a dipole phase.
We argue and give some numerical
indication that the unbinding of the
monopole-antimonopole pairs along
the time direction triggers a 2+1
dimensional unbinding, leading
to a globally Debye-screening phase.
The paper is organized as
follows: section \ref{section2} defines
the statistical problem
of the monopole plasma. The next section
focusses on the 0+1
dimensional problem along a time direction,
giving strong arguments in
favor of a phase transition.
The extension to the 2+1 dimensional
situation is then discussed,
leading to a phase diagram as a function
of $N$ and fermion filling,
which is the main result of the paper.
The conclusion is dedicated to
a comparison with previous work
and stresses open questions.
\section{Statistical mechanics of the monopole plasma}
\label{section2}
As already stressed in the introduction,
we shall assume a
Lagrangian of the form
\begin{eqnarray}
\label{Eq1}
L &=& \sum_i \sum_{\sigma=1}^{N}
\left( \overline{c}_{i \sigma}(\tau)
\frac{\partial}{\partial \tau} c_{i \sigma}(\tau) +
\overline{b}_{i \sigma}(\tau)
\frac{\partial}{\partial \tau}
b_{i \sigma}(\tau) \right) - t_f
\sum_{\langle i,j \rangle}
\left( e^{-i a_{i j}}
c_{i \sigma}^{+} c_{j \sigma} + \mbox{h.c.} \right)\\
\nonumber
&& - t_b \sum_{\langle i,j \rangle}
\left( e^{-i a_{i j}} b_i^{+} b_j
+ \mbox{h.c.} \right)
+ i \sum_j \lambda_j \left(
c_{j \sigma}^{+} c_{j \sigma} +
b_j^{+} b_j - \frac{N}{2} \right).
\end{eqnarray}
The fields $c_{i \sigma}(\tau)$ and
$b_i(\tau)$ are respectively
fermionic and bosonic, and they are
defined on a two dimensional
square lattice with continuous imaginary time.
The hopping constants $t_f$ and $t_b$
 can be derived from a large $N$
saddle point approximation of the
one band t-J model \cite{Ref5}
\cite{Ref8}. We shall from now on focus
on the effective dynamics
of the $U(1)$ gauge field $(a_{i j},
\lambda_i)$, assuming that
fermions and bosons have been traced out.
 As shown in the references
\cite{Ref3} \cite{Ref6}, the gauge field
action to gaussian order
is dominated by the fermion contribution
at low doping, and with
the assumptions that the holons have not condensed.
Keeping only the transverse part which is
 responsible of the non-Fermi
liquid behavior gives \cite{Ref3}
\begin{equation}
\label{Eq2}
S_{eff}(a,\lambda) = T \sum_{\omega_n=2 \pi n T} \int_{BZ}
\frac{1}{2} \left( \epsilon_1(k,\omega)
 \omega^{2} + \mu(k,\omega) k^{2}
\right) \left( \delta_{i,j} - \frac{k_i k_j}{k^{2}}
a_i(k,\omega) a_j(-k,-\omega) \right)
{}.
\end{equation}
In this equation,
\begin{equation}
\epsilon_1 = \frac{k_f}{2 \pi k |\omega|}
\end{equation}
for $|\omega| \ll 2 t_f k_f k$, $k
\ll k_f$ and $\mu(k,\omega)=
t_f/12 \pi$.
The gauge field variables $a_{r,r+n}$,
where $n$ is a lattice
vector and $r$ a lattice site are
denoted in the continuum limit
by $a_n(r+n/2)$, in order to define
the two component field
$a_i(\rho)$. The time component of $a$
is identified with $\lambda$.
Most of the time, we shall use the axial gauge $a_0=0$.
Latin indices such as $i$ and $j$
denote spatial components, whereas
Greek indices correspond to arbitrary components.
The quantities $\epsilon_1$ and $\mu$ are
derived with the approximation
of a circular Fermi surface, and by taking
the long wavelength,
small frequency limit of the fermion
current-current correlation function.
The main drawback of this action is that
the fundamental periodicity
of the original action (\ref{Eq1}),
namely its invariance under
$a_{i j} \rightarrow a_{i j}+2 \pi$ is lost.
This periodicity allows for non-trivial
space-time configurations
of the field corresponding to tunneling
events where the flux
threading a given plaquette may change
by integer multiples of
$2 \pi$. Ioffe and Larkin suggest to
express (\ref{Eq2}) in terms of
gauge invariant field strength
$F_{\mu \nu} = \partial_{\mu} a_{\nu}
- \partial_{\nu} a_{\mu}$, with $\mu=0,1,2$,
and to replace
the flux par plaquette $F_{1 2}$ by
its value modulo $2 \pi$.
This algorithm sounds quite natural
on physical grounds.
However, (\ref{Eq2}) has been derived
perturbatively for `flat'
configurations which satisfy Faraday's law:
$\partial_{\mu} b_{\mu}=0$, where
\begin{equation}
b_{\mu} = \frac{1}{2} \epsilon_{\mu \nu \rho} F_{\nu \rho}
{}.
\end{equation}
After the field strength $b_{\mu}$
 is taken modulo $2 \pi$,
it satisfies
\begin{equation}
\partial_{\mu} b_{\mu} = \sum_i 2 \pi n_i \delta(r-r_i)
\delta(\tau-\tau_i)
,
\end{equation}
where $n_i$ are the integer charges
located at $r_i$,$\tau_i$.
An ambiguity arises in extending the
result (\ref{Eq2}) to non
trivial configurations. We may add to
equation (\ref{Eq2}) any quadratic
form
\begin{equation}
\label{Eq3}
T \sum_{\omega_n} \int \frac{d^{2}k}
{(2 \pi)^{2}} C(k,\omega) \left(
\omega^{2} b(k,\omega) b(-k,-\omega)
- k^{2} e_{\perp}(k,\omega)
e_{\perp}(-k,-\omega) \right)
\end{equation}
without changing the result on `flat'
configurations, but the action
for non trivial configurations will
depend on the kernel $C(k,\omega)$.
In (\ref{Eq3}), $e_{\perp}$ and $b$
denote the transverse part of the
electric field, and the magnetic
field respectively.
We also note that the perturbative
evaluation of the fermion loop
generates only the function
$\epsilon_1(k,\omega) \omega^{2}
+ \mu(k,\omega) k^{2}$. Physical intuition suggests that
$\mu(k,\omega)$ is identical to the static
diamagnetic susceptibility
in the $\omega \rightarrow 0$ limit.
This determines the two functions
 $\epsilon_1(k,\omega)$ and
$\mu(k,\omega)$ as given above, and
with such a determination,
(\ref{Eq2}) becomes
\begin{equation}
S_{eff}(a,\lambda) = T \sum_{\omega_n}
\int \frac{d^{2}k}{(2 \pi)^{2}}
\frac{1}{2} \epsilon_1(k,\omega) e_{\perp}(k,\omega)
e_{\perp}(-k,-\omega) + \frac{1}{2}
\mu(k,\omega) b(k,\omega)
b(-k,-\omega)
{}.
\label{Eq4}
\end{equation}
Equation (\ref{Eq4}) is then extended to
non-trivial configurations
thus lifting the ambiguity in the choice
of the kernel $C(k,\omega)$.
But if the procedure seems perfectly sound
at low frequencies, the
separation between the electric and
magnetic parts is less obvious to
access at higher frequencies. In the
bulk of this paper, we assume
that this procedure is valid. The action for a
 many monopole configuration
with a topological charge $q(r,\tau) =
\pm 2 \pi$ is then given by
\begin{equation}
S_{plasma} = \frac{T}{2} \sum_{\omega_n} \int
\frac{d^{2}k}{(2 \pi)^{2}}
q(k,\omega) \frac{\epsilon_1(k,\omega)
\mu(k,\omega)}
{\epsilon_1(k,\omega) \omega^{2} +
\mu(k,\omega) k^{2}}
q(-k,-\omega)
\label{Eq5}
,
\end{equation}
where $q(k,\omega)$ is the Fourier transform
of the charge density, namely
\begin{equation}
q(k,\omega) = \int_0^{\beta} d \tau
\sum_r e^{-i(k.r+\omega \tau)}
q(r,\tau)
,
\end{equation}
where $r$ is a lattice site. More specifically,
this gives
\begin{equation}
S_{plasma} = \frac{N t_f}{24 \pi}
T \sum_{\omega_n}
\int \frac{d^{2}k}{(2 \pi)^{2}}
\frac{q(k,\omega) q(-k,-\omega)}
{|\omega| \left( |\omega|+
\frac{t_f}{6 k_f} k^{3} \right)}
\label{Eq6}
{}.
\end{equation}
In this formula, the global factor $N$ has
been added. It is simply
the number of fermion colors in the large $N$ approaches.
Rescaling energies and frequencies by
setting $t_f = \pm 1$,
the model depends only on two dimensionaless
parameters, $N$ and $k_f$.
Assuming a circular Fermi surface, the maximal
value of $k_f$ corresponds
to $1/2$ electron per site for a given color, which
gives $k_f \le (2 \pi)^{1/2}$.

Before going further, we should
mention that equation (\ref{Eq5}) is
not the only candidate for the
monopole plasma action.
Developing an analogy with Josephson
junction arrays, and emphasizing
the dissipative nature of the gauge
field, Nagaosa has also considered
the following action \cite{Ref10}:
\begin{equation}
S_{diss} = \frac{\gamma}{4 \pi}
\int_{-\infty}^{+\infty} d \tau
\int_0^{\beta} d \tau' \sum_{r,n}
\frac{1}{(\tau-\tau')^{2}}
\left( 1 - \cos{\left( a_{\perp}
(r,n,\tau) - a_{\perp}(r,
n,\tau') \right)}\right)
{}.
\end{equation}
for the dissipative part of the gauge field dynamics.
As shown in
reference \cite{Ref11}, it is possible to map it into
a statistical model
in some regimes, but mostly a bidimensional model is
 obtained.
The key variables are the winding number $m(r,n)$
along the time direction:
\begin{equation}
a_{\perp}(r,n,\beta) - a_{\perp}(r,n,0) =
2 \pi m(r,n) + \nu(r,n)
,
\end{equation}
with $\nu(r,n) \in ]-\pi,\pi]$ and $m(r,n)$ and integer.
It seems that both approaches respect the $2 \pi$
 periodicity and the
quadratic expansion of the gauge field around $a=0$.
In the absence of a fully first
principle derivation, we shall adopt
equation (\ref{Eq5}) as a working
hypothesis, and hope to clarify this
issue in a future work.

Going back to equation (\ref{Eq6}), it is
important to stress that for any $k$
value, the $\omega$ integral diverges if
$\lim_{\omega \rightarrow 0}
q(k,\omega)$ is non vanishing.
Therefore, we shall impose a constraint
on the allowed topological
charge configurations, namely
that $q(k,\omega=0)=0$ for any $k$.
In real space, it means that
\begin{equation}
\int_0^{\beta} d\tau q(r,\tau)=0
\end{equation}
for any plaquette located at $r$. The partition
function of the plasma
is then
\begin{equation}
Z = \sum_{n=0}^{+\infty} \frac{1}{(n!)^{2}}
\prod_{i=1}^{2 n} \left( \int_0^{\beta}
\frac{d\tau_i}{\tau_0}
\sum_{r_i} \right) \chi(r_1,...,r_n)
\exp{\left( - \frac{1}{2} \sum_{i,j}
 q_i q_j V(r_i-r_j,\tau_i-\tau_j)
\right)}
\label{Eq7}
{}.
\end{equation}
In this equation, $r_i,\tau_i$ denote
 the space-time coordinates
of the monopoles with topological charge $q_i$.
We set $q_i=2 \pi$ if $1\le i \le n$ and
$q_i=-2 \pi$ if $n+1 \le i \le 2 n$.
$\chi_{2n}(r_1,...,r_{2n})$ expresses
the constraint and $\chi=1$
if for any $r$ we have
\begin{equation}
\sum_{i=1}^{2n} q_i \delta_{r,r_i}=0
{}.
\end{equation}
The interaction potential $v(r,\tau)$ is
obtained by Fourier transform
of equation (\ref{Eq6})
\begin{equation}
v(r,\tau) = \frac{N}{12 \pi} T\sum_{\omega_n} \int
\frac{d^{2}k}{(2\pi)^{2}} \frac{e^{i(k.r+\omega \tau)}}
{|\omega| \left(|\omega| + \gamma |k|^{3} \right)}
\label{Eq8}
,
\end{equation}
with $\gamma = 1/6 k_f$ (we set $t_f=1$).
An important ingredient in (\ref{Eq7})
is the imaginary time scale
$\tau_0$ which is obtained by calculating
the ratio of the two
gaussian determinants in the presence and
in the absence of an
instanton. We have carried out this
calculation for the broken
parabola model of Ioffe and Larkin,
which leads to
\begin{equation}
\frac{1}{\tau_0} = \sqrt{2 \pi} \mu
\left( T\sum_{\omega_n}\int\frac{d^{2}k}{(2 \pi)^{2}}
\frac{k^{2}}{\epsilon_1(k,\omega) \omega_n^{2}
+ \mu k^{2}} \right)^{1/2}
{}.
\end{equation}
The interested reader will find a
derivation of this result in the
appendix. We note that the integral
is divergent at large frequencies.
This may be another signal that we
have not yet found a satisfactory
derivation of this monopole plasma action.
Since this $\tau_0$ depends on the
full non linear action of the gauge field,
which is still unknown, we shall
assume it equal to unity in the
following discussion (since we have
used $t_f=1$ as energy unit).
The following sections are now
dedicated to an analysis of the
classical statistical system
given by equation (\ref{Eq7}).

\section{Monopoles in dimension 0+1}
We deduce from the interaction (\ref{Eq8}) that
the interaction betwen two monopoles is
\begin{equation}
-V(r,\tau) = \frac{N t}{12 \pi}
\int \frac{d \omega}{2 \pi}
\int \frac{d^{2} k}{(2 \pi)^{2}}
\frac{\cos{(k.r)} -
\cos{(k.r + \omega \tau)}}
{
|\omega|\left( |\omega| + \gamma |k|^{3} \right)}
\label{eq1}
{}.
\end{equation}
We have shifted the interaction by
an infinite constant
so that the pair interaction between
two monopoles is finite.
The energy of a configuration of
monopoles satisfying the
neutrality condition $\int q(r,\tau)
 d\tau = 0$ is finite
and does not depend on the regularization
of the pair potential.

{}From now on, we are interested in the quantum problem
at zero temperature, so the system is
infinite along the imaginary
time direction. We shall now use
$\beta=Nt/12\pi$ to denote the
inverse fictitious temperature of
the monopole plasma, which shouldn't
introduce confusion.
The two parameters associated to (\ref{eq1})
are $\gamma=1/6 k_F$ and the prefactor $\beta$.
The interaction (\ref{eq1})
decreases as the distance $|{r}|$
between two plaquettes increases.
In order to have an idea of
the interaction ranges, we calculate
the interaction $V(r,\tau)$
as a function of $\tau$ for different
values of the interplaquette
distance. We first rearrange the
expression (\ref{eq1}) using
the change of variables $u = \omega \tau$ and
$q = (\gamma \tau)^{1/3} k$.
We obtain
\begin{equation}
\label{eq11}
- \frac{1}{\beta} V(r,\tau) =
\frac{\tau^{1/3}}{\gamma^{2/3}}
F \left( \frac{|r|}{(\gamma \tau)^{1/3}} \right)
,
\end{equation}
with
\begin{equation}
\label{eq12}
F(x) = \int_0^{+ \infty} \frac{q dq}{2 \pi} J_0(q x)
\int_{- \infty}^{+ \infty}
\frac{du}{2 \pi}
\frac{(1 - \cos{u})} {|u|(|u|+q^{3})}
{}.
\end{equation}
We plotted on figure \ref{fig1}
the interaction for different
values of the interplaquette distance $|r|$.
In a first approximation, we
take into account only the
one-plaquette interaction along
the time direction.
In section \ref{bidim}, we renormalize
the bidimensional problem with
a cut-off for the distance between
two plaquettes.

The interactions of the one-plaquette
problem are simply
$V(\tau) = -\tau^{1/3} F(0)/\gamma^{2/3}$.
We look for the phase diagram
of the potential $V(\tau) \sim -\tau^{\alpha}$
in one dimension as a
function of the exponent $\alpha$. Fortunately,
some exact results
concerning the phase diagram of one-dimensional
systems with
long-range interactions are available \cite{Ref12}.
The main result shows rigorously the existence
of a finite temperature
phase transition for the 1D ferromagnetic
Ising model if the coupling
$J(n-n') \propto 1/|n-n'|^{\gamma}$,
with $1<\gamma<2$ \cite{Ref13}.
To some extent, these results can be
transposed to generalized
Coulomb gas models, by considering a
representation of the Ising model
in terms of kink and antikink configurations.
The potential energy
for a single kink-antikink pair is then
proportional to
$|n-n'|^{2-\gamma}$, $n$ and $n'$ being
the locations of the
kink and antikink. The Ising and
corresponding Coulomb gas problems
are however not equivalent since the
Ising model generates only
rather special configurations where
kinks and antikinks alternate.
We expect intuitively the unrestricted
Coulomb gas to be less ordered
than the corresponding Ising model. By
ordered state, we mean the
dipolar phase. As a result, an unrestricted
generalized Coulomb gas with
$V(r) \propto r^{\alpha}$ is expected to
have a high temperature
Debye-screening phase if $\alpha < 1$.
The fact that $\alpha=1$
(the 1D genuine Coulomb potential) is
the borderline is confirmed by
several exact investigations \cite{Ref14}
\cite{Ref15} showing that
this system is always in the dipolar
phase at any temperature.
Our problem is a special case, with $\alpha=1/3$.
We shall now attempt to estimate
the transition temperature to the
Debye phase. It is then tempting
to use a real space RG analysis
along the lines of references
\cite{Ref16}-\cite{Ref18}.
For instance, the Coulomb potential
in any dimension $d$ ($\alpha=
2-d$) has been analyzed in reference \cite{Ref18}.
For $d>2$, the system is always
in a Debye screening phase, whereas
for $d<2$, there exists a finite
temperature transition. We note
that this simple RG analysis still
predicts a non trivial fixed point
for $d=1$ and $\alpha=1$, in
discrepancy with the exact results
of references \cite{Ref14} and \cite{Ref15}.
But as $d$ is decreased from 2 to 1,
the unstable fixed point is
found for higher values of the plasma fugacity,
so that the dilute approximation
leading to the RG equations
is no longer valid. Hopefully,
$\alpha=1/3$ is not too large,
so the usual RG procedure is consistent.

In order to analyse the one plaquette problem,
we wish to treat
the more general problem of the generalized
Coulomb potential
$V_{\alpha}$ in one dimension.
We show that if $\alpha<1$,
the plasma has a Debye phase.
We call $Z_{\tau}$ the partition
function of the plasma with
a minimal separation $\tau$ betwen the charges,
which position is allowed to vary from $x=0$ to $x=L$. For
a neutral system of $2 n$ particles,
this defines an integration domain
denoted $D_{2n}(L,d\tau)$.
We wish to perform
one renormalization step, that is to
express $Z_{\tau}$ as a function
of $Z_{\tau+\delta \tau}$.
To do so, we write $Z_{\tau}$ under the form
\begin{eqnarray}
\label{eq2}
Z_{\tau} &=& \sum_{m=0}^{+ \infty}
\frac{K^{2m}}{(m!)^{2}}
\sum_{p=0}^{m} {m \choose p}^{2}
p! (d \tau)^{p}
\int_{D_{2(m-p)}(L,\tau+d \tau)}
dr_1 ... dr_{2(m-p)}
W_B(r_1,...,r_{2(m-p)})\\
&&\prod_{i=1}^{p} \int_0^{L}
d \rho_i 2 \cosh{\left(
\beta q \tau E(\rho_i) \right)}
\nonumber
{}.
\end{eqnarray}
We have introduced a fugacity
denoted by $z=K \tau$.
In this equation, we have taken
into account $p$ dipoles with their
center of gravity located at $\rho_i$
($i=1,...,p$) and with a size
between $\tau$ and $\tau+d \tau$.
$W_B$ is the Boltzmann weight of the
$2(m-p)$ remaining
isolated charges located at $r_1,...,r_{2(m-p)}$.
$E(\rho)$ is the electric field
at $\rho$, created by the $2(m-p)$ isolated charges.
More specifically,
\begin{equation}
W_B(r_1,...,r_{2n}) = \exp{\left(-\beta^{-1} \sum_{i<j}
q_i q_j V_{\alpha,\tau}(r_i-r_j)\right)}
,
\end{equation}
with
\begin{equation}
V_{\alpha,\tau}(r) = - \frac{1}{\alpha}
\left( \left|\frac{r}{\tau}
\right|^{\alpha}-1\right)
,
\end{equation}
and
\begin{equation}
E(\rho) = - \sum_i q_i \nabla_{\rho}
V_{\alpha,\tau}(\rho-r_i)
= \sum_i \frac{q_i}{\tau} \left|
\frac{\rho-r_i}{\tau}\right|^{\alpha-1}
{}.
\end{equation}
The $p$ dipoles are assumed to be independent.
The integration over
the dipole coordinates $\rho_i$ must take
into account the position
of the other charges located at
$r_1,...,r_{2(m-p)}$.
We expand the $\cosh$ in (\ref{eq2})
up to second order
in the electric field, and we write
\begin{equation}
\int_0^{L} 2 \cosh{\left( \beta q
\tau E(\rho) \right)} d\rho
= 2L + \varphi(r_1,...,r_{2 n})
\label{eq4}
,
\end{equation}
where we have used the notation $n=m-p$ and
where the integration domain
takes into account the presence of
a hard core condition.
We first need to determine the function $\varphi$.
To do so, we write
\begin{eqnarray}
\int_0^{L} 2 \cosh{\left( \beta q
\tau E(\rho) \right)} d\rho
&=& \sum_{i=1}^{2 n}
\theta(x_{i+1}-x_i-3 \tau)\\
&&\int_{x_i+2 \tau/2}^{x_{i+1}-3 \tau/2} d \rho
\left( 2 + \beta^{2} \tau^{2} q^{2} \sum_{j=1}^{2n}
\sum_{k=1}^{2n} q_j q_k V'_{\tau}(\rho-x_j)
V'_{\tau}(\rho-x_k) \right)
\label{eq3}
\nonumber
{}.
\end{eqnarray}
If we take only the two-body interactions,
and the thermodynamic limit,
the expression of $\varphi$ takes the form
\begin{eqnarray}
\label{Eq10}
\varphi(r_1,...,r_{2n}) &=& - 3 \tau (2n)
+ \lim_{L\rightarrow +
\infty} \beta^{2} q^{2} \left[
\sum_{i\ne j}q_i q_j
\int_0^{L} d\rho \left|\frac{
\rho-r_i}{\tau}\right|^{\alpha-1}
\left|\frac{\rho-r_j}{\tau}
\right|^{\alpha-1} \right.\\
\nonumber
&& \left.
\mbox{sign}{\left( (\rho-r_i)
(\rho-r_j)\right)}
\theta{\left( \left|\frac{\rho-r_i}
{\tau}\right| -\frac{3}{2} \right)}
\theta{\left( \left|\frac{\rho-r_j}
{\tau}\right| -\frac{3}{2} \right)}
\right.\\
\nonumber
&&
\left. + \sum_i q^{2} \int_0^{L} d\rho
\left|\frac{\rho-r_i}{\tau}\right|^{2
(\alpha-1)}
\theta{\left( \left|\frac{\rho-r_i}
{\tau}\right| -\frac{3}{2} \right)}
\right]
{}.
\end{eqnarray}
The $L \rightarrow +\infty$ limit exists
provided the system is
neutral and $\alpha<3/2$. Indeed, the
$2n$ charges create a dipolar field
at large distances which decays as
$\rho^{\alpha-2}$ or faster.
Taking the square gives the upper
bound on $\alpha$ for long
distance convergency. This property
also enables us to shift variables
and recast the previous expression as
a sum of pair contributions which all
converge separately. We thus obtain
\begin{eqnarray}
\varphi(r_1,...,r_{2n}) &=& - 3
\tau (2n) + \beta^{2} q^{2} \sum_{i\ne j}
q_i q_j \int_{-\infty}^{+\infty}
d\rho \left[
\left|\frac{\rho-r_i}{\tau}
\right|^{\alpha-1}
\left|\frac{\rho-r_j}{\tau}
\right|^{\alpha-1} \right.\\
\nonumber
&& \left.
\mbox{sign}{\left( (\rho-r_i)
(\rho-r_j)\right)}
\theta{\left( \left|\frac{\rho-r_i}
{\tau}\right| -\frac{3}{2} \right)}
\theta{\left( \left|\frac{\rho-r_j}
{\tau}\right| -\frac{3}{2} \right)}
\right. \\
\nonumber
&& \left.
- \left| \frac{\rho}{\tau} -
\frac{r_i+r_j}{2 \tau}\right|^{2(\alpha-1)}
\theta\left( \left|\rho-\frac{r_i+r_j}
{2}\right|-\frac{3}{2} \tau \right)
\right]
{}.
\end{eqnarray}
After some rather simple calculations,
and extracting the dominant
behavior, we have
\begin{equation}
\varphi(r_1,...,r_{2n}) =
\beta^{2} q^{2} \tau \sum_{i\ne j} q_i q_j
c(\alpha) \left[ \left|
\frac{r_i-r_j}{\tau}\right|^{2\alpha-1}-1\right]
- 2 n \tau \left(3+(c(\alpha)+
d(\alpha)) \beta^{2} q^{2} \right)
{}.
\end{equation}
The coefficients $c(\alpha)$ and $d(\alpha)$
are given in terms of the
Euler $B$ function:
\begin{eqnarray}
c(\alpha) &=& \frac{2(\alpha-1)}{2\alpha-1}
B(\alpha,2(1-\alpha))-B(\alpha,\alpha)\\
d(\alpha) &=& \frac{2}{2\alpha-1}
\left(\frac{3}{2}\right)^{2\alpha -1}
{}.
\end{eqnarray}
We note that $\varphi$ has the
dimension of a length, so that $Z_{\tau}$
is dimensionless. The scaling
equations are now obtained and read
\begin{eqnarray}
\frac{d \ln{\beta}}{d \ln{\tau}} &=&
\alpha+2(2 \alpha-1) c(\alpha)
K^{2}\tau^{2}\beta q^{2}\\
\label{Eq9}
\frac{d \ln{K}}{d \ln{\tau}} &=&
- \frac{\beta q^{2}}{2} - \left(
3 + (c(\alpha)+d(\alpha))
\beta^{2} q^{4} \right) K^{2} \tau^{2}
{}.
\end{eqnarray}
And the interaction function is modified by
\begin{equation}
\frac{dV}{d \ln{\tau}} =
2(2\alpha-1) c(\alpha) K^{2} \tau^{2}
\beta q^{2} \left[ \frac{1}{2
\alpha-1} \left( \left|\frac{\Delta}{\tau}
\right|^{2\alpha-1}-1\right) -
\frac{1}{\alpha} \left(\left|
\frac{\Delta}{\tau}\right|^
{\alpha}-1\right)\right]
,
\end{equation}
where $\Delta$ is the particle separation.
These equations are obtained by imposing
the normalization constraints
$V(\tau)=0$ and
\begin{equation}
\frac{dV}{d \Delta}(\Delta=\tau) =-1
{}.
\end{equation}
Since the fundamental form of the
interactions is preserved only
for the Coulomb potential ($\alpha=1$),
these prescriptions are meaningful
mostly near $\alpha=1$.
In the case $\alpha=1$, we recover
Kosterlitz's RG equations as
derived in \cite{Ref18}. We note that
the second term in the r.h.s.
of equation (\ref{Eq9}) is not given in \cite{Ref18}, but
it doesn't change the critical behavior.
Its meaning is a natural reduction
of fugacity because of
excluded volume effects.
The structure of these equations,
and in particular, the fact that $\alpha>0$ and
$2(2\alpha-1)c(\alpha)
K^{2} \tau^{2} \beta q^{2} <0$ shows that
the model keeps a finite temperature
transition, and that the
size of the Debye screening phase
increases as $\alpha$ decreases.
Coming back to our problem, $\alpha$
is fixed to $1/3$ for the
single plaquette problem.
The interaction strength is larger if
$N$ increases and if $\gamma$ decreases,
so if $k_f$ increases.
The dipolar phase is then expected at
large $N$ and large electron
filling, in agreement with the physical
intuition that spin-charge
separation is more likely to occur in
the vicinity of the
Mott insulator and in the large $N$ limit.
This defines a critical
Fermi wave vector $k_f^{*}(N)$ such that
spin-charge separation
occurs for $k_f>k_F^{*}(N)$. The aim of section \ref{bidim}
is to obtain quantitative results for
the variation of $k_f^{*}(N)$
with the number of colors $N$.
\section{Monopoles in dimension 2+1}
\label{bidim}
We now consider the 2+1 dimensional problem.
The derivation of the RG equations
is a straightforward generalization
of what has already been done in
the 0+1 dimensional case.
The potential $V(r,\tau)$ is given
by equation (\ref{eq11})
and (\ref{eq12}). We use periodic
boundary conditions in time,
with a period $L$. This regularization
will also be used in
the numerical calculations.
Small dipoles (with a length between $\tau$ and
$\tau+d\tau$) can only be
parallel to the temporal direction
since we require the local neutrality
condition $\int q(r,\tau) d\tau=0$,
so that the cut-off $\tau$ is only
introduced in the
temporal direction.
The function $\varphi$ is given by
\begin{eqnarray}
\nonumber
\varphi(r_1 t_1,...,r_{2n} t_{2n})
&=& 2 \beta^{2} q^{2} \tau^{2}
\sum_{r} \sum_{j\ne k}
q_j q_k \int_{3 \tau/2}^{L/2}
d\tau \partial_0 V(0,\tau)
\partial_0 V(r_k-r_j-r,
\Delta_{j,k}-\tau)\\
&&+ 2 \beta^{2} q^{2} \tau^{2}
\sum_{r} \sum_j q_j^{2}
\int_{3 \tau/2}^{L/2} \left(
\partial_0 V(r,\tau) \right)^{2} d\tau
\label{eq13}
{}.
\end{eqnarray}
In this expression, the summation over
$r$ is a summation over
the plaquettes which contain the small
dipoles (with a separation in the
temporal direction between $\tau$ and
$\tau+d\tau$). $\Delta_{j,k}$
is the difference between the time
coordinates of the monopole $j$
and the monopole $k$.
As in the expression (\ref{Eq10}),
the integration over the
time coordinate of the small dipole
contains a hard core condition.
We evaluated numerically the integrals
in (\ref{eq13}), and
took into account only the potentials
$V(r,\tau)$ such as
$|r|<\Lambda$, with $\Lambda$ a lattice cut-off.
The RG trajectories are plotted on
figure (\ref{fig5}) for
different values of the Fermi wave vector.
Notice that
some trajectories are free to cross
each other since the potential
depends on the initial conditions via $\gamma$.
We can check the validity of the
predictions of the 0+1 dimensional
approach: if $k_f<k_f^{*}$, the
plasma is deconfined
whereas it is confined if $k_f>k_f^{*}$.
For a one-color model,
we find $k_f^{*} = 0.5 \pm 0.05$.
However, the Fermi wave vector
is bounded above by $\sqrt{2 \pi}$
since there is less than 1/2
electron of a given color per plaquette.
We conclude that there exists
a transition even for the one-color model.
We now adress the question of
the $N$ colors monopole model.
The action is simply multiplied by $N$,
inducing a change in the
initial conditions of the
renormalization procedure.
We plotted on figure \ref{fig6}
the critical Fermi wave vector
$k_f^{*}$ as a function of the
number of colors.
We see that for $N=2$, which is
the case of physical interest, that
a possibility of a non Fermi liquid
arises as the doping increases.

\section{Conclusion}
To conclude, the main result of this
investigation is the possibility
of tuning the microscopic parameters
of the model (here the number of
colors and the filling factor) in such
a way that confinement between
spinons and antiholons arises or not,
depending on these parameters.
Using a different approach, it had been
previously claimed that the
Ioffe-Larkin type of plasma relevant
to the t-J model is always in
the dipolar phase, so that spinons
and holons have a chance not to
form a Fermi liquid \cite{Ref10}.
It is true that our plasma action
(equation (\ref{Eq6})) is obtained
from the simplest fermion loop,
without dressing the fermion Green's
function, and this may be the source
of the difference between the results.
We were guided here by the very strong
anisotropy of the intermonopole
potential, making it much stronger
along the time direction, and
requiring the neutrality constraint
for each plaquette along the time
direction. Our intuition is that screening
may only be more effective
if the two spatial dimensions are added
to the one plaquette problem,
thus weakening the strength of the large
distance inter-monopole
interaction. We hope that these ideas
may lead to a more rigourous approach,
and possibly Monte-Carlo studies of this plasma.
Moreover, we have also presented
arguments showing that a satisfactory
microscopic derivation of the
plasma action is still missing.
One difficulty is connected to
the ambiguity present while implementing
the necessary periodicity requirement
from a perturbative
calculation which by essence assumes
`flat' field configurations.
A second one, and maybe related to
the previous remark,
is the diverging bare fugacity which
results from the Villain-type
treatment developed in the Appendix.

B.D. would like to thank J. Wheatley
and D. Khveshchenko for their
stimulating participation in earlier
attempts to address the
question of collective effects in the
monopole plasma, and S.
Sachdev for an interesting discussion.

\section{Appendix: derivation of the
monopole fugacity in the
Ioffe-Larkin approach}
We start from the quadratic action in
the transverse subspace
\begin{equation}
\label{eqA2}
S = \frac{1}{2} \frac{T}{N_s} \sum_{\omega}
\sum_k
\epsilon_1(k,\omega) e_{\perp}(k,\omega)
e_{\perp}(-k,-\omega)
+ \mu b(k,\omega) b(-k,-\omega)
{}.
\end{equation}
In this expression, $b$ is a scalar
field and $e_{\perp}$ is
related to $b$ by the Faraday equation
\begin{equation}
\omega b(k,\omega) = k \times e_{\perp}
(k,\omega).\hat{z}
{}.
\end{equation}
Suppose we consider an instanton located
on a given plaquette $r_0$
at time $\tau_0$. The idea is to replace
$b(r,\tau)$ by
$b(r,\tau)-2 \pi \theta(b(r,\tau)-\pi)
\delta_{r,r_0}$.
Minimizing over transverse configurations
of $b$ leads to the instanton
profile
\begin{eqnarray}
e_{\perp}(k,\omega) &=& -i \frac{\mu
\hat{z} \times k}
{\epsilon_1(k,\omega) \omega^{2} + \mu
k^{2}} q(k,\omega)\\
b(k,\omega) &=& \left(-\frac{i}{\omega}
+ i \frac{\epsilon_1(k,
\omega) \omega}{\epsilon_1(k,\omega)
\omega^{2} + \mu k^{2}} \right)
q(k,\omega)
\label{eqA1}
{}.
\end{eqnarray}
Here, $q(k,\omega) = 2 \pi
\exp{(-i(k.r_0+\omega \tau_0))}$ is the
corresponding topological charge density.
The fugacity is obtained from
considering quadratic fluctuations
around the single-instanton solution
and integrating them out.
We have to single out the zero mode
which corresponds to a global
translation along the time direction
of this solution.
This is a standard procedure, and we
just quote the result \cite{Ref19}
\begin{equation}
\label{eqA4}
K = \frac{A}{\sqrt{2 \pi}} \left(
\frac{\prod \epsilon_i^{(0)}}
{\prod_{i\ne 0} \epsilon_i} \right)^{1/2}
{}.
\end{equation}
In this formula, $A$ is the norm of
the zero mode function
\begin{equation}
(r,\tau) \mapsto \frac{\partial
b(r,\tau)}{\partial \tau}
{}.
\end{equation}
{}From equation (\ref{eqA1}), we obtain
\begin{equation}
A = 2 \pi \mu \left( \frac{T}{N_s}
\sum_{k,\omega}
\frac{k^{4}}{(\epsilon_1(k,\omega)
\omega^{2} + \mu k^{2})^{2}}\right)^{1/2}
{}.
\end{equation}
The second term is related to the
ratio of the product of eigenvalues
of the Hessian matrix in the vacuum
and in the presence of the instanton.
In the denominator, it is necessary
to exclude the zero eigenvalue coming
from translation invariance.
The Hessian matrix is found by expanding
the action (\ref{eqA2}) with the shift
$b(r,\tau) \rightarrow b(r,\tau)
- 2 \pi \theta(b(r,\tau)-\pi)
,$
up to quadratic order in field
deviations $\delta b(r,\tau)$
around the instanton solution.
Written in Fourier space, the
quadratic part of the action is
\begin{eqnarray}
\label{eqA3}
\delta^{2}S &=& \frac{1}{2}
\frac{T}{N_s} \sum_{k,\omega}
\Delta(k,\omega) \delta b(k,\omega)
\delta b(-k,-\omega)\\
&&- \frac{\pi \mu}{|b_0'|}
\frac{T^{2}}{N_s^{2}} \sum_{k,\omega}
\sum_{k',\omega'} e^{i((k-k')r_0+
(\omega-\omega')\tau_0)}
\delta b(k,\omega) \delta b(-k',-\omega')
\nonumber
{}.
\end{eqnarray}
We have used the notations
$\Delta(k,\omega) = \mu +
\epsilon_1(k,\omega) \omega^{2}/k^{2}$, and
\begin{equation}
b_0' = \frac{\partial b}
{\partial \tau}(r_0,\tau_0)
,
\end{equation}
where $b$ is the instanton profile.
The eigenmodes corresponding to
equation (\ref{eqA3}) are obtained
from a rational secular equation
since the scattering potential
is separable. This equation reads
\begin{equation}
- \frac{2 \pi \mu}{|b_0'|} \frac{T}
{N_s} \sum_{k,\omega}
\frac{1}{\epsilon-\Delta(k,\omega)} =1
{}.
\end{equation}
{}From the structure of this equation,
it is possible to derive the
determinant ratio as
\begin{equation}
\frac{\prod_{i\ne 0} \epsilon_i}
{\prod \epsilon_i^{(0)} } =
\frac{2 \pi \mu}{|b_0'|} \frac{T}{N_s}
\sum_{k,\omega} \frac{k^{4}}{(
\epsilon_1(k,\omega) \omega^{2}
+ \mu k^{2})^{2}}
{}.
\end{equation}
Using equation (\ref{eqA4}), and
the expression for $A$ leads to
$K=(\mu |b_0'|)^{1/2}$. From equation
(\ref{eqA1}), we finally get
\begin{equation}
K = \sqrt{2 \pi} \mu \left(
\frac{T}{N_s} \sum_{k,\omega}
\frac{k^{2}}{\epsilon_1(k,\omega) \omega^{2} + \mu k^{2}}
\right)^{1/2}
{}.
\end{equation}

\newpage
\renewcommand\textfraction{0}
\renewcommand
\floatpagefraction{0}
\noindent {\bf Figure captions}

\begin{figure}[h]
\caption{}
\label{fig1}
Interaction potential $V(|r|,\tau)$ of equation (\ref{eq1})
for $k_f = 4$ and for
different values of the interplaquette spacing as
a function of the time coordinate. The notation
(0,0) corresponds to $|{\bf r}|=0$, (1,0) stands for
$|{\bf r}|=1$, (1,1) stands for $|{\bf r}|=\sqrt{2}$
and (2,0) stands for $|{\bf r}|=2$.
\end{figure}

\begin{figure}[h]
\caption{}
\label{fig5}
RG trajectories for $N=1$ and
different values of
the Fermi wave vector. The time
coordinate is compactified on a circle
of length 400. We took $\tau=1$.
Some trajectories cross each other.
This is due to the fact that the
potential (\ref{eq11}) (\ref{eq12})
depends explicitely on $\gamma$
and thus on the initial conditions.
We have plotted the square of
the fugacity fugacity
$z^{2}=K^{2}\tau^{2}$ as a function of the effective
inverse fictitious temperature
of the monopole plasma. The dipolar
phase corresponds to the fixed
point $(\beta,z)=(\infty,0)$
and the Debye screening phase
corresponds to the fixed point
$(\beta,z)=(0,\infty)$.
\end{figure}

\begin{figure}[h]
\caption{}
\label{fig6}
Critical Fermi wave vector as a
function of the number of colors.
The errorbars indicate the precision
in the location of the
fixed point. The dashed line indicates the maximal value of
the Fermi wave vector ($\sqrt{2 \pi}$).
\end{figure}
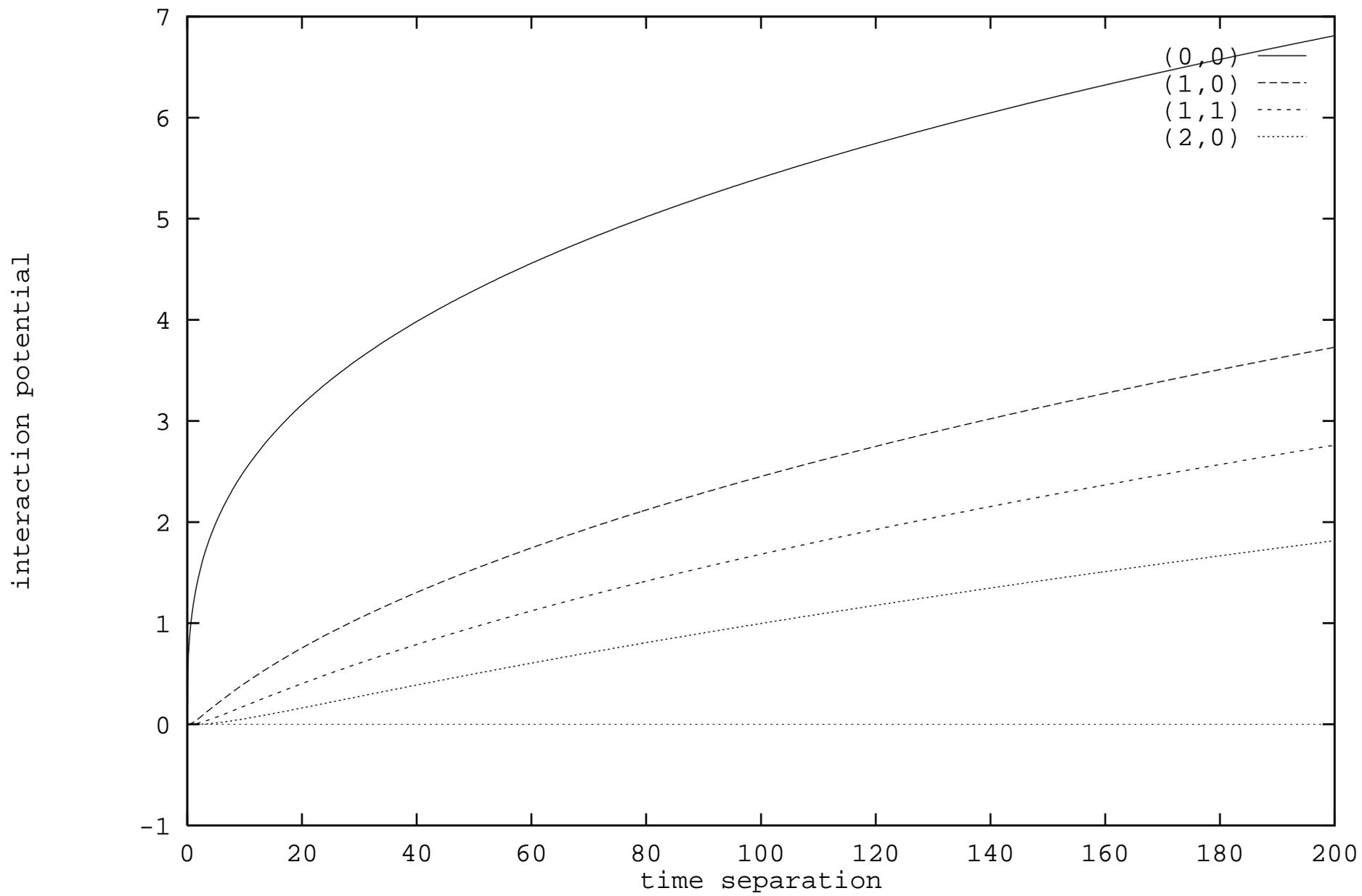
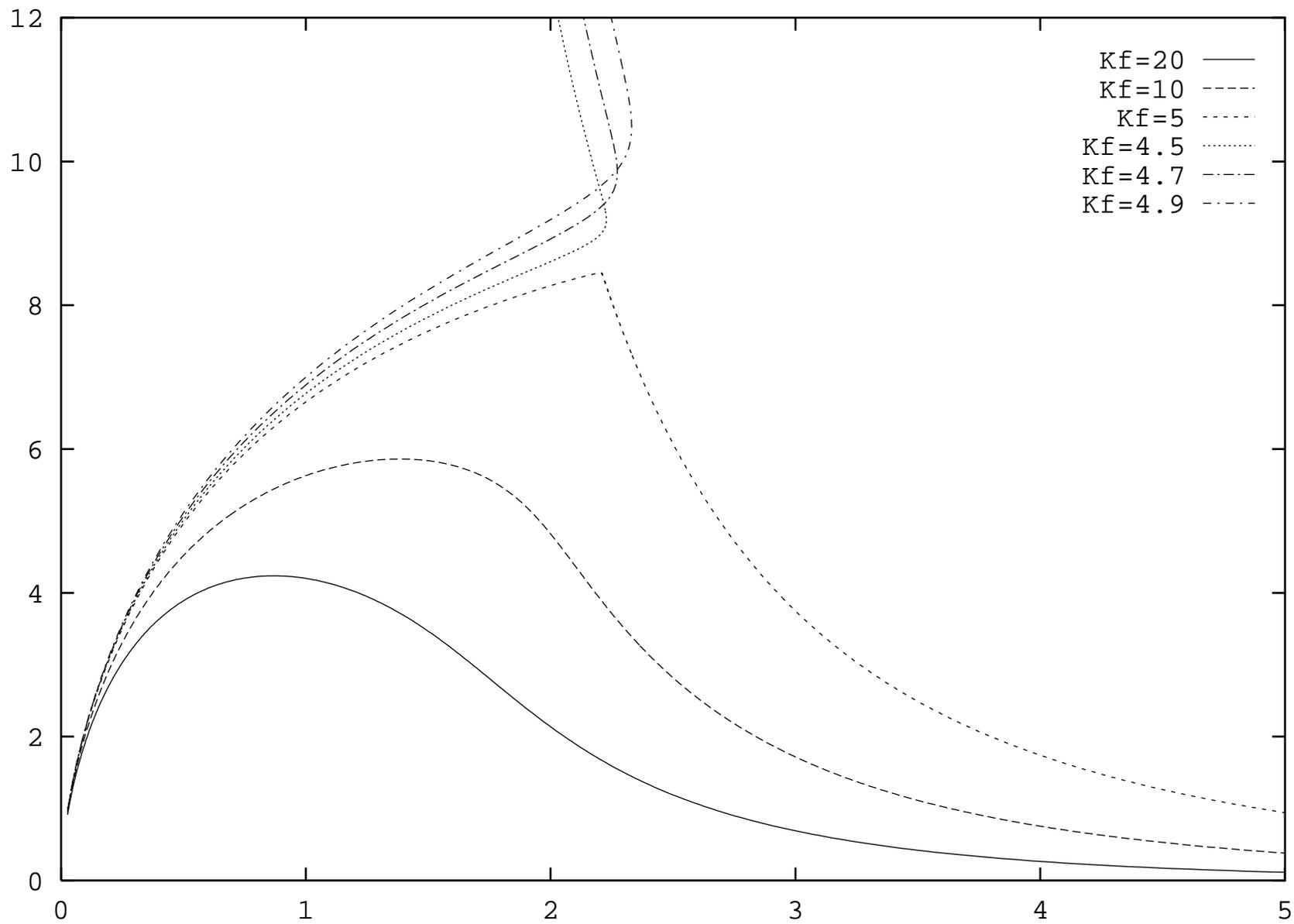
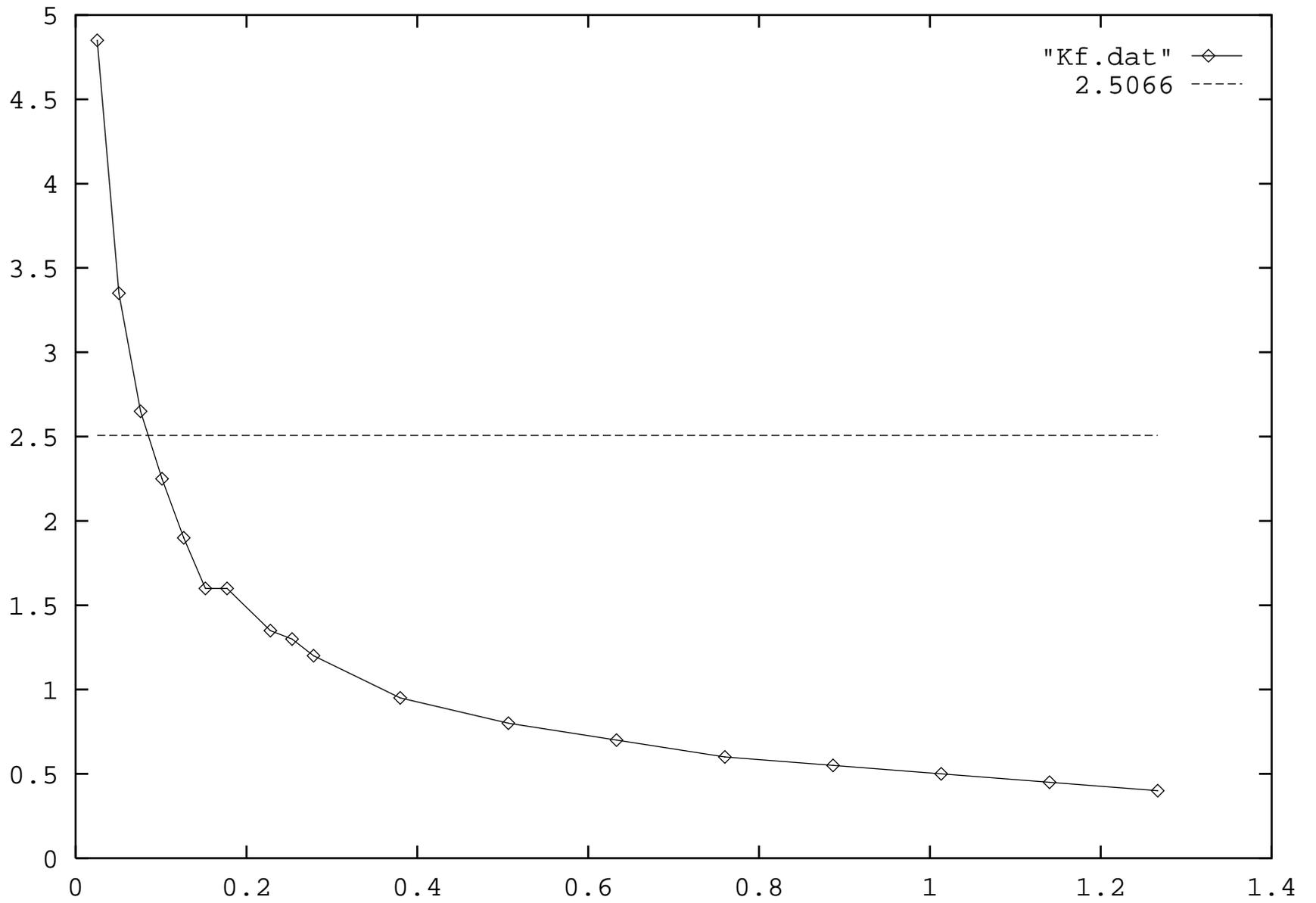
\end{document}